\documentclass[prb,twocolumn,amsmath,amssymb,floatfix,aps,showkeys,showpacs]{revtex4}

\usepackage{graphicx}
\usepackage{dcolumn}
\usepackage{bm}

\usepackage{ulem}

\begin{document}

\title{Critical behavior of the Random-Field Ising model at and beyond the Upper Critical Dimension}

\author{Bj\"orn Ahrens}
\email{bjoern.ahrens@uni-oldenburg.de}
\author{Alexander K. Hartmann}
\email{a.hartmann@uni-oldenburg.de}
\affiliation{Institute of Physics, University of Oldenburg, 26111 Oldenburg, 
Germany}

\date{\today}

\begin{abstract}
The disorder-driven phase transition of the RFIM is observed using
exact ground-state computer simulations for hyper cubic lattices in
$d=5,6,7$ dimensions. Finite-size scaling analyses are used to calculate the critical point and the critical exponents of the specific heat,  magnetization, susceptibility and of the correlation length. For dimensions $d=6,7$ which are larger or equal to the assumed upper critical dimension, $d_u=6$, mean-field behaviour is found, i.e. $\alpha=0,\;\beta=1/2,\;\gamma=1,\;\nu=1/2$. For the analysis of the numerical data, it appears to be necessary to include recently proposed corrections to scaling at and beyond the upper critical dimension.
\end{abstract}


\maketitle
\section{\label{Introduction}Introduction}
The random-field Ising model (RFIM) \cite{ImryMa1975}
is a prototypical model for magnetic systems exhibiting quenched disorder. For $d \leq 3$ and higher dimensions \cite{bricmont1987}, the model exhibits a phase transition: For low temperatures and/or weak disorder  the coupling dominates and the system is long range ordered. Increasing the temperature and/or the disorder to a critical value, the RFIM with a Gaussian distribution of the disorder
undergoes a second order phase transition to a paramagnetic phase. At  dimensions below the lower critical
dimension $d\leq d_l=2$ the critical disorder strength is zero and no ferromagnetic long-range order\ \cite{ImryMa1975,Binder1983}
occurs for finite values of the randomness. In $d=3$ the RFIM is believed \cite{fishman1979} to describe the behavior of experimentally accessible antiferromagnets well.
To investigate $d=3,4$ dimensional RFIM systems, many studies were published, see, e.g., Refs. \cite{GofmanAdlerAharonyHarrisSchwartz1993,Rieger1995,Nowak1998,art_uli1999, HartmannYoung2001,middleton2002,frontera2002,seppala2002,Hartmann2002, middleton2002b,fes_rfim2008}
and critical exponents for the magnetization $\beta$ and the susceptibility $\gamma$ are known with good accuracy.  Results for the
specific heat exponent $\alpha$ are diverse, varying from quite negative to small positive values. As a result some values support the scaling relation $\alpha + 2\beta + \gamma =2$. Also a different scaling relation $\alpha + 2\beta + \gamma =1$ has been proposed \cite{Nowak1998}. 

The RFIM has been studied by mean-field techniques as well \cite{SchneiderPytte1977,Aharony1978,Tasaki1989}.  The upper critical dimension $d_u$, above which the RFIM shows dimension-independent mean-field behavior, has been predicted  to be $d_u=6$\ \cite{ImryMa1975}. The critical exponents are believed to hold the mean field values of the pure Ising model. To the knowledge
of the authors, no numerical simulations have been performed to confirm the value of the upper critical dimension of the RFIM or to study its critical behavior close to or at $d_u$. Also corrections to scaling are still unknown in this region. In this paper we present our numerical results of the critical exponents around the upper critical dimension in $d=5,6,7$. Therefore exact ground states (GS) of large instances are calculated using a mapping to a combinatorial optimization problem. The main result is to confirm $d_u=6$ and all exponents for $d_u\ge 6$ to be indeed compatible with the mean-field values of the pure Ising ferromagnet, if corrections to scaling are taken into account. A summary of this work can be found at the {\tt papercore} database \cite{papercore}.

The RFIM consists of a hyper cubic lattice of Ising spins $S_i=\pm 1$. The Hamiltonian of the RFIM is given by
\begin{equation}
\mathcal{H}=-J\sum_{\langle i,j\rangle}S_iS_j - \sum_i\left(h_i +H\right)S_i\,.
\end{equation}
$J$ denotes the ferromagnetic coupling constant between two adjacent spins. $\langle i,j\rangle$ denotes to sum over next
neighboured spins only. Each spin is exposed to a net random-field $h_i+H$ which consists of two contributions: $h_i=h\varepsilon_i$ is the quenched local random-field with a disorder strength $h$. The $\varepsilon_i$ are distributed according to a Gaussian with zero mean and unity width. $H$ indicates the strength of a homogeneous magnetic field, which is used to determine the susceptibility, but is set to $H=0$ elsewhere.

The paper is organized as follows: In the next section we briefly outline the numerical approach we have used. Also we state in detail all quantities we measured and the finite-size scaling (FSS) approaches we have used to analyze the data. In section three we present our main results. In the last section, we discuss our work and conclude with a summary.

\section{\label{MM}Method and measured quantities}
\begin{figure}[!ht]
\includegraphics[width=0.4\textwidth]{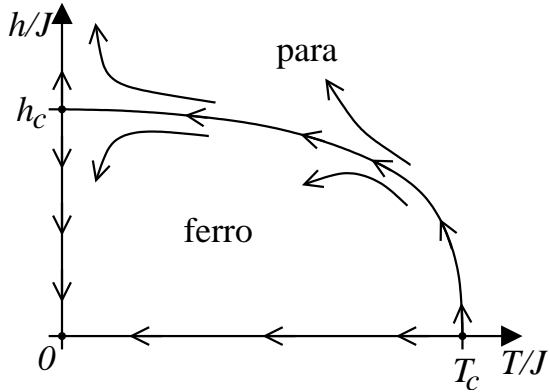}
\caption{Schematic phase diagram and renormalization group flow of the RFIM.
\label{fig:phase-diagram}}
\end{figure}

From mean-field theory, the phase transition of the Gaussian RFIM is known to be second order along the whole phase boundary\
\cite{SchneiderPytte1977}. Furthermore, renormalisation group (RNG) theory brought up the existence of three fixed points for the
renormalization group flow in the temperature/disorder phase space of the RFIM\ \cite{BrayMoore1985}. There exists a stable fixed point at $T=h=0$, an unstable fixed point at $T=T_c,\,h=0$ and a saddle-point at $T=0,\,h=h_c$, see Fig.\ \ref{fig:phase-diagram}. 
 The saddle-point is stable against changes of the temperature and unstable to pertubations of the disorder $h$. A crossing of the phase boundary at finite temperatures ($T>0$) leads to the same critical behaviour as a crossing at $T=0$ by varying $h$ does.
This allows us to focus on ground states only, which is very convenient, especially from the computational point of view. We
calculate the GS by mapping each realization $(\{h_i\},H)$ of the net random-field to a graph with suitable edge capacities. For this graph we apply a sophisticated maximum flow/minimum cut algorithm known from algorithmic graph theory to calculate the minimum cut\ \cite{PicardRatliff1975,ogielski1986,GoldbergTarjan1988,HartmannRieger2001}.  

The cut separates the graph into two parts. They directly correspond to the GS spin configuration $\{S_i\}$ of that specific realization of the net disorder. This approach allows us to obtain exact ground states of large systems with a linear lattice lengths up to $L=14$ in $d=5$, $L=9$ in $d=6$ and $L=8$ in $d=7$. In contrast, Monte Carlo would need an unresonable amount of CPU time for the same data quality, due to long equilibration times.

\begin{figure}[!ht]
\includegraphics[width=0.4\textwidth]{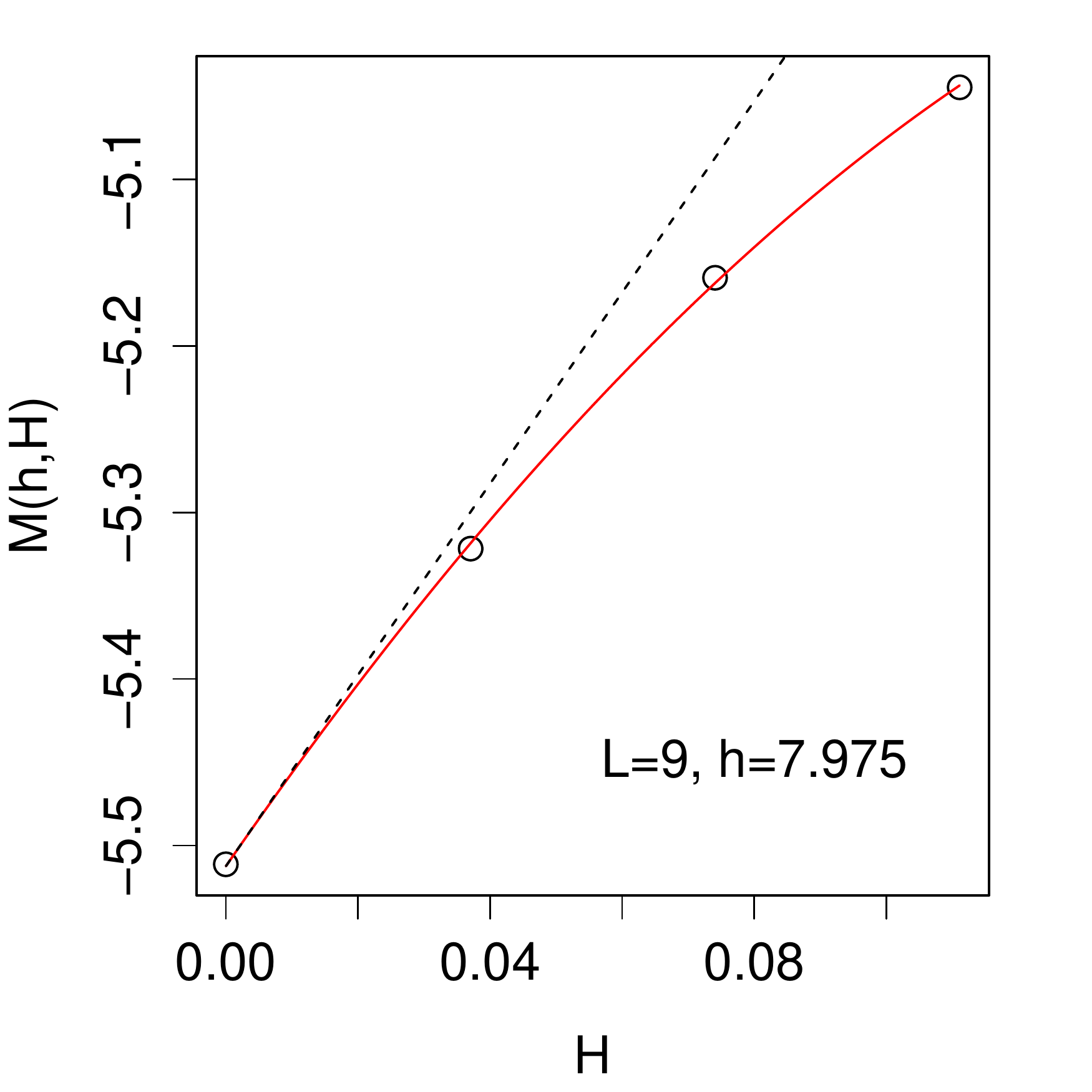}
\caption{(color online) Approximation of the susceptibility in $d=6$ by fitting a parabola to different homogeneous magnetic fields using $H_1=0.036$. For this realization we obtain via the coefficient of the linear term $\chi_{\text{single}}\approx5.2$.\label{GetSusceptibility}}
\centering
\end{figure}

From a GS spin configuration, we calculated directly the quantities of interest, which are magnetization per spin ($N$ is the number of spins)
\begin{equation}
 M=\frac{1}{N}\sum_i^N S_i\;
\end{equation}
and the bond energy per spin
\begin{equation}
E_J=\frac{J}{N}\sum_{\langle i, j \rangle} S_i S_j\;.
\end{equation}
From these individual values, we obtained averaged quantities like the average magnetization $m=[M]_h$, where $\left[\cdots \right]_h$  denotes the disorder average for a given value of $h$. We also calculated the zero-temperature susceptibility
\begin{equation}
\chi(h)=\left [ \left. \frac{\partial M(\{h_i\},H)}{\partial H}\right|_{H=0} \right]_h\;,\label{SuszDef}
\end{equation}
The actual value of the susceptibility is obtained as linear response of the single-disorder realization magnetization to small homogeneous magnetic fields $H$. Therefore, we apply small homogeneous fields at equidistant values $H_1,2H_1,3H_1$ to each realization of the disorder and (automatically) fit parabolas as function of $H$ to the obtained single realization magnetizations $M(H=0,h)$,  $M(H=H_1,h)$,  $M(H=2H_1,h)$ and  $M(H=3H_1,h)$. The average of all linear coefficients of the parabolas at a particular $h$ then is the susceptibility $\chi(h)$ (see Figure\ \ref{GetSusceptibility}). $H_1$ is chosen depending on the dimension and the size of the system, see Sec.\ \ref{Results}.

We also calculate a specific heat like quantity $C(h)$ as numerical derivative of the bond energy per spin $E_J$ (see \cite{HartmannYoung2001} for details) which is defined as
\begin{equation}
 C(h)= \frac{\left [\partial E_J(h)\right]_h}{\partial h}\;.\label{SpecificHeat}
\end{equation}
From the magnetization at $H=0$ we deduce the Binder cumulant\ \cite{Binder1981},
\begin{equation}
g(h,L) = \frac{1}{2}  \left(  3 - \frac{  \left[  M^4  \right]_h }{  \left[ M^2  \right] ^2_h}   \right).
\label{eq:Binder}
\end{equation}

We are interested in the infinite-size limit and the critical behaviour of these quantities, especially at the critical point. Our analysis proceeded in the following way: For each quantity we start at $d=5$, where standard FSS arguments according to\ \cite{PrivmanFisher1984} are valid, continuing with $d=7$, in which a correction-to-scalilng theory exists. After that we take care of its scaling in $d=6$. We started at the "easy" quantities, as the Binder parameter and proceeded with the magnetization, susceptibility and at last the specific heat.

In $d=5$, standard FSS says, that finite-size effects enter via the lattice length to correlation length ratio $L/\xi_\infty$. Since the correlation length at the critical point diverges as $\xi_\infty\sim |h-h_c|^{-\nu}$,  one gets $L/\xi_\infty\sim L|h-h_c|^{\nu}$. The same is valid for the Binder cumulant, $g(h,L)$. Taking the argument to the power of $1/\nu$, one expects the Binder cumulant to scale as
\begin{eqnarray}
g(h,L) & = & \tilde g\left((h-h_c)L^{1/\nu}\right) 
\;\;\;(d < d_u) ,\label{binder_form} \\
& \equiv & \tilde g(X^{(d)}(L,h)) \nonumber
\end{eqnarray}
Equality is a result of the dimensionless definition of the Binder cumulant. Using a properly chosen  $h$-axis rescaling function $X^{(d)}(L,h)$, the Binder cumulants for all system sizes belonging to the same dimension collapse on one single (unknown) master curve. Hence, finding the critical exponent $\nu$ is now left as a problem with one degree of freedom: it is the value of $\nu$ where the collapse of the rescaled data points from different system sizes onto one curve is best.
 
The argument of the rescaling function in Eq.\ (\ref{binder_form}) vanishes at the critical point $h=h_c$. Thus the curves of the Binder cumulant do intersect.  But for $d\geq d_u$ strong corrections to scaling arise and the point of intersection for different system sizes drifts systematically. This phenomenum was discussed by Jones and Young\ \cite{JonesYoung2005} for the pure Ising model. They found that for $d\geq d_u$  the lattice length has to be replaced by a larger one,
i.e.,
\begin{equation}
 L\rightarrow \tilde L=\begin{cases}
					a_6 L\ln^{1/d_u} L 	& d=6\\
					L^{{d}/{d_u}}			& d>d_u
                       \end{cases}\label{LRescaling}
\end{equation}
Thus, the scaling exponents of the correlation length depend on the dimension of the considered setup. In $d=7$ it is $\nu \rightarrow \nu^\text{MF}\!\!\cdot\!{d_u}\!/\!{d} =0.5\!\cdot\!{6}\!/\!{7}\approx 0.4286$. For $d=d_u$, the standard finite-size scaling arguments have to be adapted. Finding a correct form for $X^{(d_u)}(L,h)$ was a tedious task. The first approaches to find the scaling form at the upper critical dimension began in the early 1980's, when Br\'ezin studied the $\phi^4$-model ($d_u=4$) on hypercubic $d$-dimensional lattices\ \cite{Brezin1982}. Using RNG calculations he found multiplicative logarithmic corrections to scaling at the upper critical dimension.

Much later, in 1997 Luijten and Bl\"ote \ \cite{LuijtenBloete1997} published a scaling form for $\mathcal{O}(n)$ spin models with tunable ferromagnetic long-range interactions and tunable upper critical dimension. Their RNG calculations differ from Br\'ezin's result by 2nd order logarithmic corrections. For the Ising model with long-range interactions at $d=d_u=2$ their scaling form was numerically verified  by Gr\"uneberg and Hucht  \cite{GruenebergHucht2004}. But since the model contains no disorder we adapted some parameters and use the ansatz
\begin{equation}
X^{(6)}(h,L) = (h-h_c)L^{1/\nu_\text{MF}}\log(L)^{1/6} + b\log(L)^e \label{LuijtenBloete_scaling}
\end{equation}
 for FSS at $d=d_u=6$.

But now, let us return to the other quantities of interest and lets discuss the scaling assumptions for them. The next less easy quantity is the magnetization. From scaling theory we expect the magnetization in $d=5$ to scale as 
\begin{alignat}{1}
\label{magRescale}
m(h,L) 	&=L^{-\beta/\nu} \tilde m\left(X^{(5)}(L,h)\right)\\
		&= L^{-\beta/\nu} \tilde m\left((h-h_c)L^{1/\nu} \right)\;.
\end{alignat}
so, replacing $X^{(5)}\mapsto X^{(d)}$ and $L \mapsto \tilde L$ should result in a data collapse for the higher dimensional setups.
In the same way, a collapse of the susceptibility should be achieved, using 
\begin{alignat}{1}
\chi(h,L) &= \tilde L^{\gamma/\nu} \tilde\chi\left( X^{(d)}(\tilde L,h) \right)
\end{alignat}
For the scaling behavior of the specific heat like quantity we use
\begin{alignat}{1}
C(h,L) &= \tilde L^{-\alpha/\nu} \tilde C\left ( X^{(d)}(\tilde L,h)\right), 
\end{alignat}
since we are not aware of a suitable, analytically sound scaling form.

We observe the behaviour of those quantities as functions for different linear lattice lengths $L$ to apply finite-size scaling techniques. While the error-estimates of the observables are calculated from $200$ bootstrap samples\ \cite{Hartmann2009} the finite-size scaling analysis is done using two different tools from\ \cite{autoScale2009} and\ \cite{Huchtfsscale2.81}. The former is a command line tool based on a simplex algorithm written in python, while the latter is a hand operated rescaling plot program with direct visualisation written in C. The simplex algorithm needs about $30$ data points near the critical point to give reasonable results. As our data points for larger system sizes are relatively sparse compared to $30$ but very accurate, we created artificial data points in that region. We did this by fitting cubic splines to our data points near the critical point. The splines are used to calculate a dense set of artificial data points to reach the necessary number to serve the simplex algorithm to create a reasonable data collapse. These data points are just
auxiliary points, which means that we do not show any artificial data points in our final data-collapse plots.

\section{\label{Results}Results}
We have performed exact GS calculations for the Gaussian  RFIM in dimensions
$d=5,6$ and $7$. To perform a finite-size scaling analyses, we considered different system sizes up to $L=14$ ($d=5$), $L=9$ ($d=6$) and $L=8$ ($d=7$). This means, we obtained exact GSs for systems with up to $2\times10^6$ spins.
Each time, we averaged all results over many realizations and studied
each realization for several values of the disorder strength $h$. Note
that for the calculation of the susceptibility, each time four GSs subject to the  homogeneous fields $H=0,H_1,2H_1,3H_1$ were necessary. The homogeneous fields are chosen such that $m(H)$ stays in the parabolic domain, see also the discussion in\ \cite{HartmannYoung2001}. For details see Tab.\ \ref{tab:parameters}. Note that the values of $H_1$ for $d=5$ were chosen accidentally non-monotonously, but they are anyway small enough such that $m(H)$ is parabolic, regardless of the system size.

\begin{table}
 \centering
\begin{tabular}{ r | r l || l | r l || l | r l }
\multicolumn{3}{c}{$d=5$} & \multicolumn{3}{c}{$d=6$} 						& \multicolumn{3}{c}{$d=7$} \\
$L$  & $n_{\rm real}$ 	& $H_1$ 	& 	$L$  & $n_{\rm real}$ 	& $H_1$			& 	$L$  & $n_{\rm real}$ 	& $H_1$	\\ \hline
	&				&			&	3	&	60000		& 0.0120		&	3	&	100000		& 0.00350\\
4   	&  80000 		& 0.00200 	&	4	& 	80000		& 0.0050		&	4   	& 	80000		& 0.00150\\
5 	&160000		& 0.00075	&	5	&	30000		& 0.0025		&	5	&	5500		& 0.00075\\
6	&  39500		& 0.00075	&	6	&	15000		& 0.0015		&	6	&	2400		& 0.00035\\
7	&  72000		& 0.00050	&	7	&	7000		& 0.0010		&	7	&	8000		& 0.00025\\
8	&   40000		& 0.00030	&	8	&	3500		& 0.0006		&	8	&	5000		& 0.00010\\
9	&   10000		& 0.00100	&	9	&	1000		& 0.0004		&\multicolumn{2}{c}{}\\
10	& 	4400		& 0.00020	&\multicolumn{6}{c}{}\\
11	& 	2000		& 0.00010	&\multicolumn{6}{c}{}\\
12	& 	6800		& 0.00030	&\multicolumn{6}{c}{}\\
13	& 	3200		& 0.00030	&\multicolumn{6}{c}{}\\
14	& 	5000		& 0.00025	&\multicolumn{6}{c}{}
\end{tabular}
\caption{For each dimension $d$ and linear size $L$, the maximum
number $n_{\rm real}$ of realizations and the 
value $H_1$ of the smallest homogeneous field used to obtain the susceptibility are listed.
\label{tab:parameters}}
\end{table}

\begin{figure}[!ht]
\includegraphics[width=0.5\textwidth]{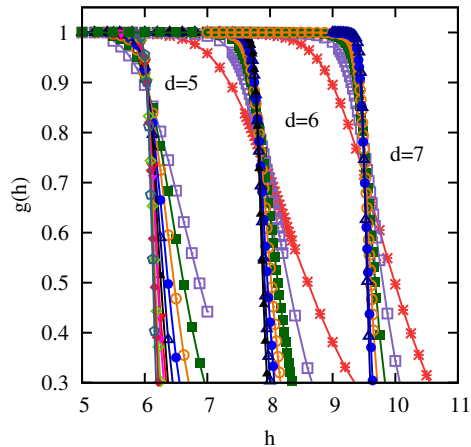}
\caption{(color online) Binder cumulant in $d=5,6,7$ dimensions for different lattice sizes $L=4\dots14$, $L=3\dots9$ and $L=3\dots8$, respectively . The lines are guides to the eyes only.\label{all_binder_cumulants}}
 \centering
\end{figure}

\begin{figure}[h]
\includegraphics[width=0.5\textwidth]{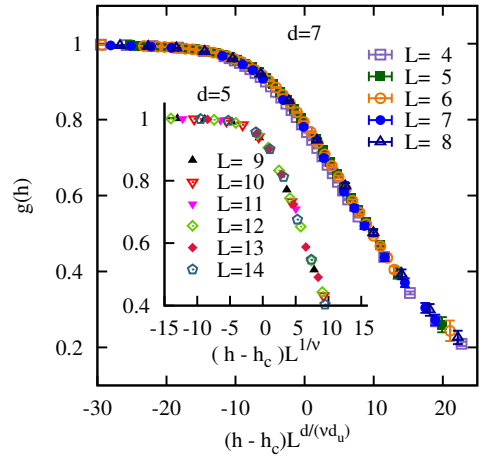}
\caption{(color online) Collapsed Binder cumulants in $d=7$ and $d=5$
(inset) for several system
sizes.\label{7d5d_binder_cumulants_collapse}   } \centering
\end{figure}

\begin{figure}[h]
\includegraphics[width=0.5\textwidth]{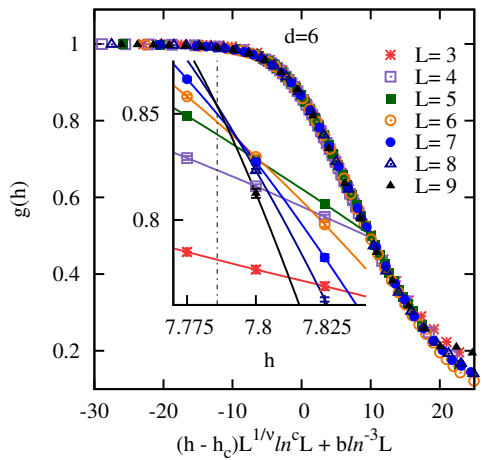}
\caption{(color online) Collapsed Binder cumulants for $d=6$ (main plot). The
 inset shows the critical region where the estimated curves of the
 Binder parameter intersect. The dotted black line corresponds to the
 value $h=7.7859$.\label{binder_collapse_d6_inset_intersections} }
 \centering
\end{figure}

To get a first impression of the nature and location of the phase transition, we start by showing results for the Binder cumulant
Eq.\ (\ref{eq:Binder}).  As mentioned above, for second-order phase transitions, the Binder cumulants obtained for different system sizes
at or close to a critical point  will intersect.  The reader may get a rough guess of the critical disorder strength from the point of intersection observing Fig.\ \ref{all_binder_cumulants}. Based on the data
of the largest system sizes, our first estimates  
are $h^{(5)}_c=6.02(1)$ in $d=5$,  $h^{(6)}_c=7.78(1)$ in $d=6$  and
 $h^{(7)}_c=9.48(5)$ in $d=7$. The intersection is rather clear for
$d=5$. Using an appropriate magnification, it
can be seen in the inset of Fig.\
\ref{binder_collapse_d6_inset_intersections}, that for
$d=6$ the points of intersection vary systematically. A similar also systematic
drift may be observed for $d=7$, but due to larger error bars, we do
not quantify them.
A more reliable way to determine the critical point is to apply finite-size scaling theory and collapse the data to a master curve. For $d=5$ we follow the scaling ansatz according to Eq.\ (\ref{binder_form}). Our data collapse best is for $h^{(5)}_c=6.0157(10)$ and $\nu^{(5)}=0.626(10)$. For $d=7$  we used $\nu \rightarrow \nu \frac{d_u}{d}$ obtaining
$h^{(7)}_c=9.4889(1)$ and $\nu^{(7)} = 0.49(2)$ which is compatible with the mean-field result $\nu_\text{MF}=0.5$. The collapsed data
curves are shown in Fig.\ \ref{7d5d_binder_cumulants_collapse}.  The data collapses look quite good for larger system sizes at $d=5,7$. When including the smallest sizes, the quality of the data collapses decreases. For $d=7$ it improved a bit by including 
some higher order corrections obtained by trial and error. Since we have no hint on the precise form of the corrections, we do not go into details and use the results from the high-quality collapse when omitting  $L=3$.

In contrast to this, the data collapse for the Binder parameter for  $d=6$ looks only good for all system sizes when applying a
 scaling ansatz which includes logarithmic corrections according to  Eq.\ (\ref{LuijtenBloete_scaling}).  The best collapse is achieved
 using  $h^{(6)}_c = 7.7859(1)$, $\nu^{(6)}=0.51(5)$, $c = 1/6$, $b = 3.403$, and  $e = -3.0$ (see Fig.\ \ref{binder_collapse_d6_inset_intersections}).

\begin{figure}[!ht]
\includegraphics[width=0.43\textwidth]{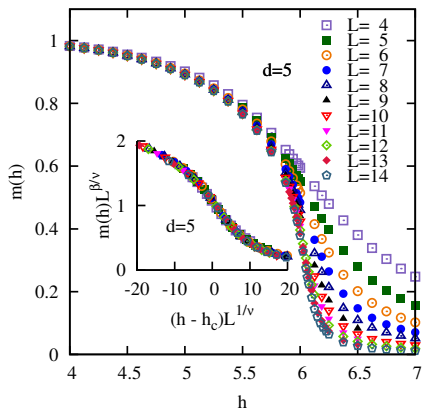}
\caption{(color online) Magnetization (main plot) and collapsed magnetisation (inset) in
 $d=5$ for several system sizes.\label{D5_magnetization_collapse} }
 \centering
\end{figure}

The rescaling of the disorder axis according to $X^{(d)}(h,L)$ should be valid for all other quantities, so we will keep it fixed from here
on. This allows us to determine the magnetization exponent $\beta$ via rescaling the magnetization values. Using the same methods as above, we obtain in five dimensions $\beta^{(5)}=0.255(10)$ for all system sizes. The collapsed data curves are shown in Fig.\
\ref{D5_magnetization_collapse}.

\begin{figure}[!ht]
\includegraphics[width=0.5\textwidth]{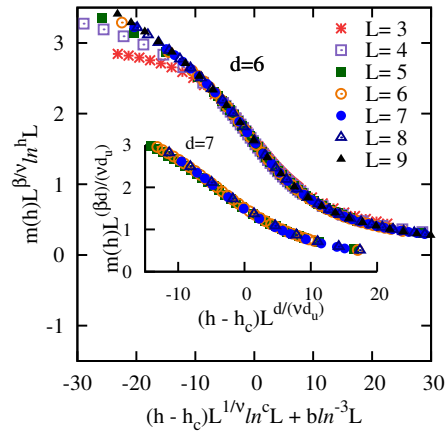}
\caption{(color online) Collapsed magnetisation in $d=6$ for several system sizes and $d=7$ (inset).\label{magnetization_collapse_d6_inset_d7}}
 \centering
\end{figure}

More interesting are the cases for $d=6,7$. Eq.\ (\ref{LRescaling}) tells us to scale the magnetization in $d=7$ with $\beta^\text{MF}\!/\!\nu^\text{MF}\,{d}\!/\!{d_u}=1.1\overline{6}$, which gives the best collapse for our data, but for the four largest
system sizes only (see inset Fig.\ \ref{magnetization_collapse_d6_inset_d7}). Due to the large deviation of the magnetization of the smaller system sizes, we tried to include some higher order terms. We achieved some collapses including these
curves, but a definite form of the corrections is unclear.

We also achieve good results in $d=6$, see Fig.\ \ref{magnetization_collapse_d6_inset_d7}, using a scaling form as
\begin{equation} 
 \tilde m (h) \sim m(h)L^{\beta/\nu}\ln^h(L),
\end{equation}
where $\beta^{(6)}\!=\!0.50(1)$ and $h=-0.33(1)$.  It is also possible to collapse the data using a pure power law,  $\tilde m (h) \sim m(h)L^{0.80(5)} \Rightarrow\beta^{*}=0.40(1) $ utilizing the same disorder rescaling to achieve the same collapse quality. Nevertheless, the resulting value $\beta^{*}$  is not compatible with the mean-field value $\beta_{\rm MF}=0.5$, see Tab.\ \ref{allResults}. This indicates that indeed also here logarithmic corrections are important. 

\begin{figure}[!ht]
\includegraphics[width=0.5\textwidth]{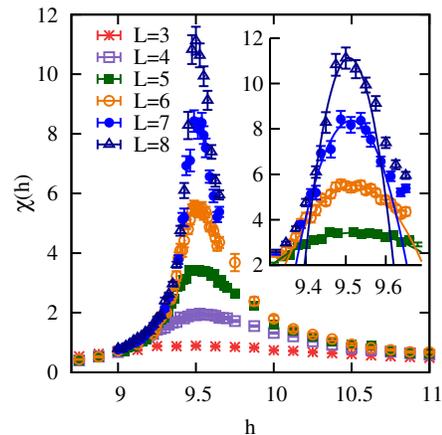}
\caption{(color online) Susceptibility in $d=7$. The inset shows the parabolas fitted to the peaks.\label{D7_susz_inset_susz_parabola_fits}}
 \centering
\end{figure}

We now turn to the susceptibility, as defined in Eq.\ (\ref{SuszDef}). It has a very clear peak at the critical point, which can be seen for example in Fig.\ \ref{D7_susz_inset_susz_parabola_fits}. Taking the peaks as a characteristic property of the susceptibility we approximate them with parabolas.

We use the maxima of the obtained parabolas for a FSS analysis. Following Jones and Young\ \cite{JonesYoung2005} we use the pure lattice length $L$ in $d=5$ and a stretched lattice length $\tilde L$ for the larger dimensions, i.e. logarithmically enlarged $\tilde L_6=a_6L\ln^{1/d_u} L$ for $d=6$ and $\tilde L_7=a_7L^{d/d_u}$ for $d=7$ and try a pure power law ansatz according to 
\begin{equation}
\chi_{\max}(\tilde L) = s_0 {\tilde L}^{\gamma/\nu}\label{susz_fss}
\end{equation}

\begin{figure}[!ht]
\includegraphics[width=0.5\textwidth]{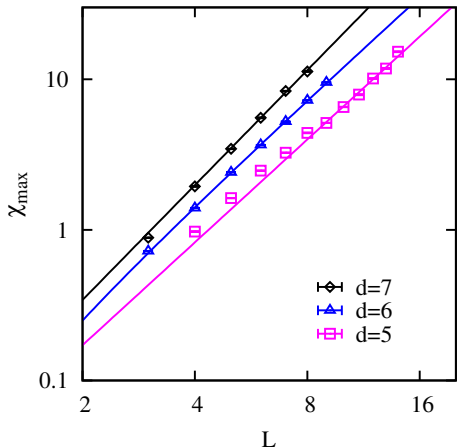}
\caption{(color online) Log-log plot of the maxima of the susceptibilities as function of the lattice length for different dimensions and their fitted power law model. The lines show the fits according to Eq.\ (\ref{susz_fss}).\label{susz_peaks_all}}
\centering
\end{figure}

For $d=5$ we leave out the smallest systems for our analysis, since they do need unknown small-system size correction terms. The results of the fits are given in Tab.\ \ref{tab:fits}. The original data points  together with the fits are shown 
in Fig.\ \ref{susz_peaks_all}. 
\begin{table}[h]
\begin{center}
\begin{tabular}{ l | c c | c }
$d$ & $s_0$ & $\gamma/\nu$ &$\gamma$\\\hline
 $5$& $0.036(9)$ & $2.27(11)$&$ 1.42(7)$\\
$6$ & $0.064(2)$ & $2.13(2)\phantom{1}$&$1.07(3)$\\
$7$ &$0.050(2)$ &  $2.24(2)\phantom{1}$ &$1.06(7)$ 
\end{tabular} 
\end{center}
\caption{Resulting fit parameters from fitting the peak height 
$\chi_{\max}$
of the  susceptibilities as a function of system size $L$ according to Eq.\  (\ref{susz_fss}).
\label{tab:fits}}
\end{table}

\begin{figure}[!ht]
\includegraphics[width=0.5\textwidth]{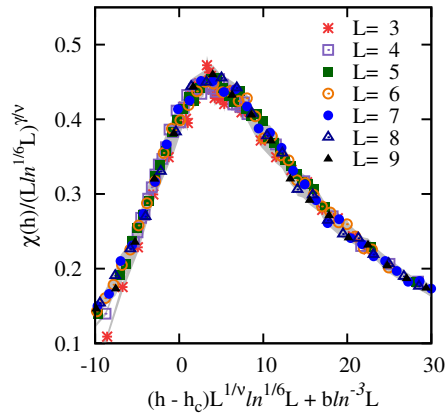}
\caption{(color online) Collapsed susceptibility  in $d=6$ for several system sizes. The gray shaded area is the cumulative error of all data points, connected through c-splines, using $\nu=0.51(5 )$, $\gamma=1.07(5)$ and $h^{(6)}_c=7.7859$}\label{D6_susz_collapse}
 \centering
\end{figure}

\begin{figure}[!ht]
\includegraphics[width=0.5\textwidth]{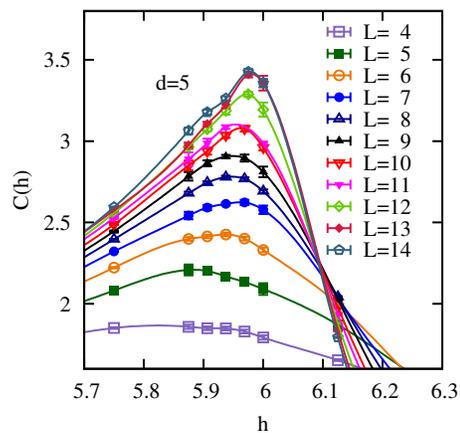}
\caption{(color online) Specific heat in $d=5$ for several system sizes. The lines are guides to the eyes only. \label{D5_specH}}
 \centering
\end{figure}

\begin{figure}[!ht]
\includegraphics[width=0.5\textwidth]{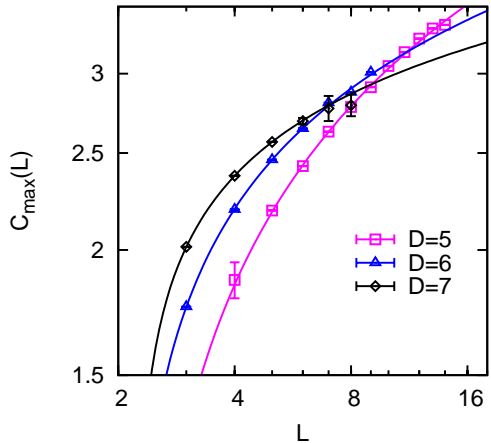}
\caption{(color online) Specific heat peaks for $d=5,6,7$ with fits according to Eq.\ (\ref{specH_peaks_fit})  \label{all_specH_peaks_logscale} on log-log scale.}
 \centering
\end{figure}

Since the values obtained for $\gamma$ are slightly larger than the mean-field values  we check the results collapsing the data. The finite-size behaviour of the disorder is taken from the results of the Binder-parameter analysis and stays unchanged from here on. The
collapse of the susceptibility of the RFIM at zero temperature at $d=6$ is shown in Fig.\ \ref{D6_susz_collapse} using the fixed value $\nu=0.51(5 )$, $\gamma=1.07(5)$ and $h^{(6)}_c=7.7859$. Given the error bars  of the data points, the data collapse appears to be fair.
The next quantity we present here is the specific heat as defined in
Eq.\ (\ref{SpecificHeat}). The curves of the specific heat for all
system sizes in $d=5$ are shown in Fig.\ \ref{D5_specH}.  One can
clearly observe a maximum, which lies slightly below $h_c$. We 
determined for all dimensions and system sizes  the positions and 
heights of the peaks by fitting parabolas to the peak locations. 
 For the very large system, as $d=6,L=9$ and $d=7,L=8$ where the
peaks of the specific heat are too ragged for a decent fit, we draw
an ellipsoid around its most likely position, taking its radii as
error bars.
From the peak heights we observe the finite-size scaling behavior.  The data points show a strong curvature in the log-log plot, 
so we first try an ansatz, including a single correction term, as 
\begin{equation}
C_{\text{max}}(\tilde L)=C_0\tilde L^{\alpha/\nu} \left(1 + a_1 \tilde L^e \right)\label{specH_peaks_fit}.
\end{equation}
with $\tilde L$ being the stretched lattice length. Finite values for the specific heat exponent $\alpha$ do not lead to proper fits. Only for $d=7$ the fit procedure converges to $\alpha/\nu=-0.0016\pm 0.089$, else it quits at $\alpha/\nu=10^{-30}$, leaving the other values unchanged. Peak heights and curves are shown in Fig.\ \ref{all_specH_peaks_logscale} using the values listed in Tab.\ \ref{tab:c-fit}.
\begin{table}
 \centering 
\begin{tabular}{l | c c c c}
$d$& $C_0$					& $\alpha/\nu$ 		&  $a_1$	 			& $e$\\ \hline 
$5$& $6.16(83)$				& $0(\text{fixed})$ 	& $-1.13(8)$ 	 			&$-0.34(9)\phantom{7}$ \\
$6$& $3.95(18)$ 				& $0(\text{fixed})$ 	& $-1.17(6)$ 	 			&$-0.67(8)\phantom{7}$ \\
$7$& $3.1(9)\phantom{77}$  	& $-0.00(9)$ 			& $-1.6(4)\phantom{7}$ 	&$-1.21(47)$
\end{tabular}.
\caption{Results of the fits of the peak-height of the specific
heat according Eq.\ (\ref{specH_peaks_fit}). \label{tab:c-fit}}
\end{table}
Since the data of the specific heat at the peaks for the larger lattices are relatively rugged and the flanks are not, we had a look at the slopes of the specific heat. Interestingly, it seems that the minimum of the slope lies right at the critical point. 
But we did not elaborate on that point.

To check our scaling assumptions we performed a data collapse.

\begin{figure}[!ht]
\includegraphics[width=0.5\textwidth]{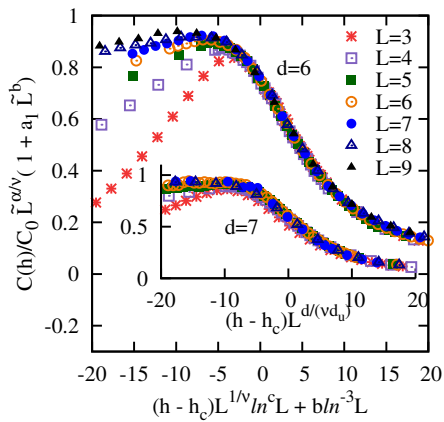}
\caption{(color online) Collapse of the specific heat in $d=6$ (main plot) using $e=-0.66$ and in the inset $d=7$ using $e=-0.23$. \label{D6_specH_collapse_inset_D7_specH}}
 \centering
\end{figure}

The obtained values lead to a good collapse for $d=5$. For $d=6,7$ the data
of the smaller system sizes do not collapse very well, but still
OK. Using slightly changed parameters gives visually a better result, as shown in Fig.\ \ref{D6_specH_collapse_inset_D7_specH}. Nevertheless, the fit concerns only the corrections to scaling, and the improvement of the fit concerns the region away from the critical point. Therefore our conclusions are not changed. 

\begin{table}
 \centering
\begin{tabular}{c  | l l l lll}
 			&  $d=3$ & $d=4$ 	& $d=5$		& $d=6$ &$d=7$ 	&MF\\ \hline
$h_c$		& $ 2.28(1)$	& $4.18(1)$	& $6.0157(10)$	& $7.7859(1)$	&$9.4889(1)$	& $\!2d\sqrt{\frac{\pi}{2}}$\\
$\beta$		& $ 0.017(5) $& $0.13(5) $	& $0.27(1)$	& $0.50(1)$		& $0.50(3)$		& $1/2$\\
$\gamma$	& $ 1.98(7)$	& $1.57(10) $	& $1.42(7)$	& $1.07(3)$		&$1.06(7)$		& $ 1 $\\
$\alpha$	& $0$  		& $ 0 $		& $ 0 $		& $ 0 $			& $0$			& $ 0 $\\
$\nu$ 		& $ 1.37(9)$	& $0.78(10)$	& $0.626(10)$	& $0.51(5)$		& $0.49(2)$		& $1/2$\\
\end{tabular}

\caption{Numerical results of the ground-state calculations for $d=34,5,6,7$
(data for $d=3$ was taken from Refs.\ \cite{HartmannYoung2001,middleton2002},
data for $d=4$ was taken from Ref.\ \cite{Hartmann2002})
and the mean-field predictions (MF) for comparison\label{allResults}.}
\end{table}

\section{\label{Conclusions}Conclusions}
We have used a mapping to the max flow problem to calculate exact GSs of the RFIM at different disorder strenghts near the critical point in $d=5,6,7$ dimensions. From those GSs we have calculated several thermodynamic quantities to determine the critical
exponents that govern the FSS behaviour of the considered observables. A summary of the values thus obtained is given in Tab.\ \ref{allResults}.

In this way, we have verified that the upper critical dimension of the RFIM is $d_u=6$ and that it exhibits mean-field critical behaviour
at $d_u$.  In particular  we obtained strong evidence for the specific heat of the RFIM to  converge towards a constant for infinite systems sizes and found $\alpha=0$.  Arguing
from the extremely good data collapses in $d=6$, a valid form of the
logarithmic corrections of the disorder finite-size scaling  for the
short range RFIM has been found and quantified.

For $d=7$ we also find mean-field exponents when rescaling the system size according to Eq.\ (\ref{LRescaling}), using a corrected correlation-length exponent for FSS analysis as $\nu^{(7)}=\nu_\text{MF}\,{d_u}\!/\!{d}$. 

The results in $d=5$ closed the gap to former simulations. Including them, a monotonic behaviour of the critical exponents
as a function of the dimension  can be seen.

Finally, since the mean-field values for the exponents hold in
$d=6,7$, the exponents do fulfill the Rushbrooke equation $\alpha +
2\beta + \gamma =2$ exactly. For $d=5$ we also get $\alpha^{(5)} +
2\beta^{(5)} + \gamma^{(5)} = 2.0$.

To summarize, our results for high dimensions together with the results 
obtained earlier for $d=3,4$  result in a rather complete picture
of the RFIM order-disorder phase transition.

\section{\label{Ack}Acknowledgments}
We like to thank Alfred Hucht for the kind and valuable discussions and Oliver Melchert for the critical reading of the manuscript. The calculations were carried out on \textbf{GOLEM} (\textbf{G}ro\ss rechner \textbf{OL}denburg f\"ur \textbf{E}xplizit \textbf{M}ultidisziplin\"are \textbf{F}orschung) at the University of Oldenburg.
\newpage
\bibliographystyle{apsrev}
\bibliography{Literatur}
\end{document}